\documentclass[12pt]{article}
\let\ssection=\section
\renewcommand{\section}{\setcounter{equation}{0}\ssection}

\newcommand\mathC{\mkern1mu\raise2.2pt\hbox{$\scriptscriptstyle|$}
        {\mkern-7mu\rm C}} 
\newcommand{\mathR}{{\rm I\! R}}         

\newcommand\eq[1]{Eq.\ (\ref{#1})}

\newcommand\ket[1]{\,|#1\rangle}
\newcommand\bra[1]{\langle #1|\,}
\newcommand\LH{L({\cal H})}
\newcommand\SP{{\rm SP}({\cal H})}
\newcommand\R{{\bf R}}

\begin{document}
\begin{titlepage}

\begin{center}
{\large\bf A Topos Perspective on State-Vector Reduction}
\end{center}

\vspace{0.8 truecm}

\begin{center}
            C.J.~Isham\footnote{email: c.isham@imperial.ac.uk}\\[10pt]
            The Blackett Laboratory\\
            Imperial College of Science, Technology \& Medicine\\
            South Kensington\\
            London SW7 2BZ\\
\end{center}
\begin{center}
            August, 2005
\end{center}

\begin{abstract}
A preliminary investigation is made of possible applications in
quantum theory of the topos formed by the collection of all
$M$-sets, where $M$ is a monoid. Earlier results on topos aspects
of quantum theory can be rederived in this way. However, the
formalism also suggests a new way of constructing a `neo-realist'
interpretation of quantum theory in which the truth values of
propositions are determined by the actions of the monoid of
strings of finite projection operators. By these means, a novel
topos perspective is gained on the concept of state-vector
reduction.
\end{abstract}
\end{titlepage}

\section{Introduction}\label{Sec:Introduction}
The goal of quantum cosmology is to describe in quantum terms the
physical universe in its entirety. As a field of study, quantum
cosmology is usually construed as a branch of quantum gravity,
although some of its most important questions transcend any
particular approach to the latter subject.

In this context, it is noteworthy that all the major approaches to
quantum gravity assume more or less the standard quantum
formalism, both in regard to its mathematical form and to its
interpretative framework. Whether such an assumption is justified
is debatable, and I have argued elsewhere that, in particular, the
{\em a priori\/} assumption of a continuum field of numbers (real
or complex) would be problematic in a theory where space and time
are not representable by a smooth manifold \cite{Isham03}. Indeed,
it may well be that the entire quantum formalism is only valid in
the atomic and nuclear realms, and that something entirely new is
needed at the scale of the Planck length.

Nevertheless, in the present paper I shall assume that the
standard mathematical formalism of quantum theory is correct and
then ask the recurrent question of whether this formalism can
yield an interpretation that lies outside the familiar
instrumentalism of the standard approach with its emphasis on
measurements made by an observer who exists `outside' the system.
That one does not wish to invoke an external observer is easy to
understand in quantum cosmology.

A simple realist philosophy would aspire to associate with each
state $\ket\psi$ a definite value for each physical quantity $A$;
equivalently, to each proposition of the form ``$A\in \Delta$''
(signifying that the physical quantity $A$ has a value that lies
in the range $\Delta$ of real numbers) there would be associated a
truth value $V^{\ket\psi}(A\in\Delta)$ that is either $1$ (true)
or $0$ (false). However, the famous Kochen-Specker theorem
\cite{KS67} prohibits the existence of any such valuation, and,
for those interested in quantum cosmology, this leads to the major
challenge of finding a interpretation of the quantum formalism
that is non-instrumentalist but which, nevertheless, does not rest
on simple `true-false' valuations.

One possible response to this challenge is to use topos theory. A
topos is a category (so there are objects, and arrows from one
object to another) with the special property that, in certain
critical respects, it behaves like the category of sets
\cite{MM92}. In particular, just as normal set theory is
intimately associated with Boolean algebra (the `Venn diagram'
algebra of subsets of a set is Boolean) so a topos is associated
with a more general algebra connected to the sub-objects of
objects in the topos.

Concomitantly, in topos theory, one encounters situations in which
propositions can be only `partly' true. The associated truth
values lie in a larger set than $\{0,1\}$, but still maintain the
distributive character of classical logic. More precisely, the
truth values in a topos lie in what is known as a `Heyting
algebra', which is a generalisation of the Boolean algebra of
classical logic: in particular, a Heyting algebra is distributive.
The main difference, however, is that, in a Heyting algebra, the
law of excluded middle may no longer hold. In other words, there
may be elements, $P$, of the logic such that $P\lor \neg P<1$
where, here, `$<$' means `strictly less than' in the partial
ordering associated with the logic. This situation is typical of
so-called `intuitionistic logic' and has been much studied by
mathematicians concerned with the formal foundations of their
subject. The important thing about a logic of this type is that it
forms a genuine deductive system---and, as such, can be used as a
foundation for mathematics itself---provided only that proof by
contradiction is not allowed.

The notion of a proposition being only `partly true', seems to fit
rather well with the fuzzy picture of reality afforded by quantum
theory, and the possibility of seriously applying topos ideas to
this subject is very intriguing. One attempt, that places much
emphasis on the use of generalised truth values, can be found in a
series of papers by the author and collaborators \cite{IB98}
\cite{IB99} \cite{HBI00}
 \cite{IB02} \cite{Ish05}. The fundamental observation in this
approach is that if we have a proposition ``$A\in\Delta$'' for
which\footnote{The quantity ${\rm Prob}(A\in\Delta;\ket\psi)$
denotes the quantum mechanical probability that the proposition
``$A\in\Delta$'' is `true' when the quantum state is $\ket\psi$.
In the standard instrumentalist interpretation(s) of quantum
theory, the proposition being `true' means that if a measurement
is made of the physical quantity $A$ then the result will
definitely be found to lie in $\Delta\subset\mathR$. For a
normalised state $\ket\psi$ we have that ${\rm
Prob}(A\in\Delta;\ket\psi)=\bra\psi\hat E[A\in\Delta]\ket\psi$
where $\hat E[A\in\Delta]$ is the spectral projector onto the
eigenspace of $\hat A$ associated with eigenvalues that lie in
$\Delta\subset\mathR$.} $0<{\rm Prob}(A\in\Delta;\ket\psi)<1$ then
although we cannot say that ``$A\in\Delta$'' is either true or
false (which would correspond to ${\rm
Prob}(A\in\Delta;\ket\psi)=1$ and ${\rm
Prob}(A\in\Delta;\ket\psi)=0$ respectively), nevertheless this
proposition may imply other propositions to which the formalism
assigns probability $1$, and which therefore {\em can\/} be said
unequivocally to be true. What is not at all obvious, but is
nevertheless the case, is that the collections of all such
propositions form a distributive logic, and therefore it is
possible to {\em define\/} the truth value of the proposition
``$A\in\Delta$'' to be the set of all propositions $P$ that are
implied by ``$A\in\Delta$'' and which are such that ${\rm
Prob}(P;\ket\psi)=1$.

In detail, there is considerably more to the idea than just this,
and in the original paper \cite{IB98}, we began by introducing the
notion of {\em coarse-graining\/} in which the proposition
``$A\in\Delta$'' is replaced by the `coarser'
proposition\footnote{The key point here is that the proposition
``$A\in\Delta$'' implies the proposition ``$f(A)\in f(\Delta)$''
although the converse is generally false. For example, if a
physical quantity $A$ has the value $2$ then this implies that the
value of $A^2$ is $4$. On the other hand, from the knowledge that
$A^2=4$ we can deduce only that $A=2$ or $-2$.} ``$f(A)\in
f(\Delta)$'' for some function\footnote{In normal set theory, the
notation $f:X\rightarrow Y$ means that $f$ is a function from the
set $X$ to the set $Y$. In a general category, the notation
$f:X\rightarrow Y$ will denote an arrow/morphism whose domain is
the object $X$ and whose range is the object
$Y$.}$f:\mathR\rightarrow\mathR$. We then ascribed to
``$A\in\Delta$'' the truth value\footnote{In general, the notation
$A:=B$ means that the quantity $A$ is {\em defined\/} by the
expression $B$. This is frequently of the form stating that $A$ is
the set of entities that possesses a particular property, as in
the example of \eq{Def:VpsiAinDeltaO}.}
\begin{equation}
    V^{\ket\psi}(A\in\Delta):=\{f_{{\cal A}({\cal H})}:\hat A\rightarrow\hat B
        \mid {\rm
    Prob}(f(A)\in f(\Delta);\ket\psi)=1\}. \label{Def:VpsiAinDeltaO}
\end{equation}
In this approach, the bounded, self-adjoint operators on $\cal H$
are viewed as the objects in a category ${\cal A}({\cal H})$, and
a function $f:\mathR\rightarrow\mathR$ defines an arrow from $\hat
A$ to $\hat B$ if $\hat B=f(\hat A)$. This is the significance of
the notation in \eq{Def:VpsiAinDeltaO} where the right hand side
is to be regarded as a `sieve'\footnote{\label{footnotesieve}A
collection $S$ of arrows with domain $O$ is said to be a  `sieve
on $O$' if for any $f\in S$, $h\circ f\in S$ for all arrows $h$
that can be combined with $f$ (i.e., are which are such that the
domain of $h$ is equal to the range of $f$). Thus a sieve is like
a {\em left ideal\/} $I$ in a monoid $M$ since $nm\in I$ for all
$n\in M$ and $m\in I$. This is one way of understanding why left
ideals in monoids are important in topos theory: something that is
much exploited in the current paper.} of arrows on the object
$\hat A$ in the category ${\cal A}({\cal H})$. One of the
fundamental results in topos theory is that, in any category, the
collection of sieves on an object form a Heyting algebra, and
hence \eq{Def:VpsiAinDeltaO} assigns (contextualised) multi-valued
truth values in quantum theory. The actual topos in this example
is given by the collection of presheaves\footnote{A `presheaf'
${\bf F}$ over a category $\cal C$ is defined to be (i) to each
object $A$ in $\cal C$, an assignment of a set ${\bf F}(A)$; and
(ii) to each arrow $f:A\rightarrow B$ in $\cal C$, an assignment
of a map ${\bf F}(f):{\bf F}(A)\rightarrow {\bf F}(B)$ such that
if $f:A\rightarrow B$ and $g:B\rightarrow C$ then ${\bf F}(g\circ
f):{\bf F}(A)\rightarrow {\bf F}(C)$ satisfies ${\bf F}(g\circ
f)={\bf F}(g)\circ{\bf F}(f)$. It is also required that if
$1_A:A\rightarrow A$ is the identity arrow at any object $A$ in
$\cal C$, then ${\bf F}(1_A):{\bf F}(A)\rightarrow {\bf F}(A)$ is
the identity map. } over the category ${\cal A}({\cal H})$.

From a mathematical perspective this structure is correct,
nevertheless the underlying theory---of presheaves and the logic
of sieves---is not the easiest thing to grasp. So it is natural to
wonder if there might be a mathematically simpler way to use topos
theory in quantum physics. For example, one can rewrite
\eq{Def:VpsiAinDeltaO} as
\begin{eqnarray}
    V^{\ket\psi}(A\in\Delta)&:=&\{f_{\cal O}:\hat A\rightarrow\hat B
        \mid {\rm Prob}(f(A)\in f(\Delta);\ket\psi)=1\}
                \label{Def:VpsiAinDeltaOO}\\[2pt]
            &=&\{f:\mathR\rightarrow\mathR\mid \hat E[f(A)\in
            f(\Delta)]\ket\psi=\ket\psi\}\label{Def:VpsiAinDelta}
\end{eqnarray}
where, in general, $\hat E[B\in\Gamma]$ denotes the spectral
projector onto the eigenspace of the (bounded, self-adjoint)
operator $\hat B$ associated with eigenvalues that lie in the
range $\Gamma\subset\mathR$.

By rewriting \eq{Def:VpsiAinDeltaO} in the form
\eq{Def:VpsiAinDelta} nothing is lost, and yet
\eq{Def:VpsiAinDelta} looks simpler since it deals directly with
functions $f:\mathR\rightarrow\mathR$, rather than with the arrows
that they induce in the category ${\cal A}({\cal H})$. In this
respect, a key observation is that the right hand side of
\eq{Def:VpsiAinDelta} is actually a {\em left
ideal\/}\footnote{Recall that a subset $I\subset M$ is a `left
ideal' if $mI:=\{mn\in M\mid n\in I\}\subset I$ for all $m\in M$.}
in the monoid of functions from $\mathR$ to $\mathR$. A left ideal
is much like a sieve of arrows (c.f.\ footnote
{\ref{footnotesieve}) and yet, arguably, is easier to grasp
intuitively.

The present paper takes its cue from replacing
\eq{Def:VpsiAinDeltaOO} with \eq{Def:VpsiAinDelta}, and is
grounded in an attempt to exploit the topos structure associated
with any monoid, not least because in text books on topos theory
this example is invariably introduced early on, and it is a
relatively easy one with which to work.

We recall that a monoid is a semi-group with an identity, and thus
differs from a group in that inverses of elements may not exist.
One obvious example of a monoid is the set of all $n\times n$
matrices in which the combination law is matrix multiplication;
the identity is then just the unit matrix. Another basic example
of a monoid is the collection, ${\rm Map}(X,X)$, of all functions
$f:X\rightarrow X$ from some set $X$ to itself, with the
combination $f\star g$ of a pair $f,g$ of such functions being
defined as their composition: $f\star g(x):=f(g(x))$ for all $x\in
X$. The monoid identity is just the identity function ${\rm
id}_X:X\rightarrow X$.

For any given monoid $M$, a key concept is that of a (left)
`$M$-set'. This is defined to be a set $X$ together with an
association to each $m\in M$ of a map $\ell_m:X\rightarrow X$ such
that (i) if $1$ denotes the unit of $M$ then $\ell_1(x)=x$ for all
$x \in X$; and (ii) for all $m,n\in M$, we have
\begin{equation}
\ell_m\circ\ell_n=\ell_{mn}\label{gammamgamman}
\end{equation}
For simplicity, the element $\ell_m(x)\in X$ will usually be
written as $mx$, and then \eq{gammamgamman} reads
\begin{equation}
    m(nx)=(mn)x         \label{m(nx)=(mn)x}
\end{equation}
for all $x\in X$.

As theoretical physicists, we are very familiar with $M$-sets for
the special case when $M$ is a group: for example, any linear
representation of a group is an $M$-set, as is the action of a
group on a manifold in the theory of non-linear group
realisations. Indeed, \eq{gammamgamman} describes a `realisation'
of the monoid $M$ in the monoid, ${\rm Map}(X,X)$, of all
functions of $X$ to itself; as such it can be viewed as a
significant generalisation of the idea of a non-linear realisation
of a group. As we shall see in the present paper, there are
potential physical roles for $M$-sets in situations where $M$ is
definitely {\em not\/} a group.

The relation to topos theory becomes clear with the observation
that, for any given monoid $M$, the collection of all $M$-sets can
be given the structure of a topos. The objects in this category
are the $M$-sets themselves, and the arrows/morphisms between a
pair of $M$-sets are the equivariant\footnote{A function
$f:X\rightarrow Y$ between $M$-sets $X$ and $Y$ is {\em
equivariant\/} if $f(mx)=mf(x)$ for all $m\in M$, $x\in X$.}
functions between them. A crucial object in any topos is the
`object of truth values', $\Omega$, which plays the analogue of
the set $\{0,1\}$ in the category of sets. In the case of the
topos of $M$-sets, $\Omega$ turns out to be the set of left ideals
in $M$. The close resemblance of a left ideal to a sieve of arrows
suggests that it might be possible to recover our earlier results
using $M$-sets rather than the more complicated mathematics of
presheaves. This is indeed the case but, as we will see, using the
theory of $M$-sets it is also possible to obtain quite new ideas
about generalised quantum valuations.

The basic mathematics of the theory of $M$-sets is described in
Section \ref{SubSec:GeneralTheory}. This is applied in Section
\ref{SubSec:MonoidClassPhys} to recover the topos ideas in
classical physics that were first discussed by Jeremy Butterfield
and myself in \cite{IB99}. Then, in Section \ref{SubSec:MonoidQT},
we show how topos monoid ideas can be used to recover in a new
guise our earlier results on quantum theory as encapsulated in
\eq{Def:VpsiAinDeltaO}. The monoid used in this example is that
given by the collection of all bounded, measurable functions from
$\mathR$ to $\mathR$.

Then, in Section \ref{Sec:ToposReduction} we strike out in a new
direction by considering possible roles for the monoid  of all
bounded operators on the Hilbert space of the quantum theory. In
turn, this leads us to consider the monoid consisting of finite
strings of projection operators and hence, finally, to a new topos
perspective on the familiar, albeit controversial, process of
state vector reduction.

\section{Monoid actions and generalised truth values}
\subsection{The general theory}\label{SubSec:GeneralTheory}
Following standard practice, we denote by $BM$ the category whose
objects are (left) $M$-sets, and whose arrows are $M$-equivariant
maps. Thus, if $X$ and $Y$ are $M$-sets, an arrow $f:X\rightarrow
Y$ in the category $BM$ is a map $f:X\rightarrow Y$ such that
$f(mx)=mf(x)$ for all $m\in M$, $x\in X$.

In any topos a key role is played by the `truth object' $\Omega$.
This object has the property that the sub-objects of any object
$X$ are in one-to-one correspondence with arrows\footnote{For the
category of sets, $\Omega$ is just the set $\{0,1\}$. If $J$ is a
subset of the set $X$ then the associated characteristic map
$\chi^J:X\rightarrow\{0,1\}$ is
\begin{equation}
        \chi^J(x):=\left\{\matrix{\mbox{$1$ if $x\in J$;\ \ }\cr
                                 \mbox{$0$ otherwise.}}\right.
\end{equation}}
$\chi:X\rightarrow\Omega$. For the category $BM$, the truth object
is the set $LM$ of all left ideals in the monoid $M$. The action
of $M$ on $LM$ is \cite{Goldblatt}
\begin{equation}
        \ell_m(I):=\{m^\prime\in M\mid m^\prime m\in I\}
                                    \label{Def:taum}
\end{equation}
for all $m\in M$. It is immediately clear that the right hand side
of \eq{Def:taum} is indeed a left ideal in $M$, and one verifies
trivially that \eq{gammamgamman} (or, equivalently,
\eq{m(nx)=(mn)x}) is satisfied. Note that for the ideal $1:=M$ we
have $\ell_m(1)=1$ for all $m\in M$. For the ideal $0:=\emptyset$,
we have $\ell_m(0)=0$ for all $m\in M$.

The Heyting algebra structure on $LM$ is defined as follows. The
logical `and' and `or' operations are $I\land J:=I\cap J$ and
$I\lor J:=I\cup J$ respectively, and the unit element and zero
element in the algebra are $1:=M$ and $0:=\emptyset$ respectively.
The partial order is defined by saying that $I\prec J$ if and only
if $I\subseteq J$, and the logical implication $I\Rightarrow J$ is
defined by \cite{Goldblatt}
\begin{equation}
        I\Rightarrow J:=\{m\in M\mid \ell_m(I)\subset \ell_m(J)\}.
\end{equation}
As in all Heyting algebras, $\neg I$ is defined by $\neg
I:=I\Rightarrow 0$; thus, in $BM$,
\begin{equation}
        \neg I:=\{m\in M\mid \forall n, nm\not\in I\}.
\end{equation}

Our task, then, is to seek physical applications for truth values
that lie in the Heyting algebra of all left ideals in a monoid.
From the perspective of topos theory, the natural way of finding
such truth values arises from the fundamental nature of
sub-objects: namely, the existence of a one-to-one correspondence
between sub-objects of an object $X$ and arrows from $X$ to
$\Omega$. In the case of a topos $BM$, the sub-objects of an
object $X$ in $BM$ are the $M$-invariant subsets of $X$, where a
subset $Y$ of $X$ is said to be `$M$-invariant' if for all $m\in
M$ and $y\in Y$ we have $my\in Y$. Then, a $BM$-arrow
$\chi:X\rightarrow LM$ (i.e., $\chi$ is an $M$-equivariant
function from $X$ to $LM$) determines the subset
\begin{equation}
    J^\chi\subset X:=\{x\in X\mid \chi(x)=1\}
\end{equation}
which, as can readily be checked, is indeed $M$-invariant.
Conversely, if $J\subset X$ is an $M$-invariant subset of $X$,
then the associated `characteristic arrow' $\chi^J:X\rightarrow
LM$ is defined by
\begin{equation}
        \chi^J(x):=\{m\in M\mid mx\in J\}. \label{Def:chiJ}
\end{equation}
It is easy to see that, since $J$ is $M$-invariant, the right hand
side of \eq{Def:chiJ} is indeed a left-ideal in $M$, and hence an
element of $LM$.

One can think of the right hand side of \eq{Def:chiJ} as being a
measure of the `extent' to which $x$ is an element of $J$: the
more elements of $M$ send $x$ into $J$ (i.e., the larger the right
hand side of \eq{Def:chiJ}) the `closer' $x$ is to being in $J$.
With this in mind, we rewrite \eq{Def:chiJ} as
\begin{equation}
    [x\in J]_{BM}:=\{m\in M\mid mx\in J\} \label{Def:xinJ}
\end{equation}
and view \eq{Def:xinJ} as the truth value in the topos $BM$ for
the proposition ``$x\in J$''. Note that if $x$ belongs to $J$ then
$[x\in J]_{BM}=M$---the unit element of the Heyting algebra $LM$.

In practice, we shall use a slight generalisation of the example
of \eq{Def:xinJ}. Namely, if $X$ is an $M$-set let ${\bf
K}:=\{K_m,m\in M\}$ be a family of subsets of $X$ that satisfy the
conditions, for all $m$,\footnote{If $K$ is any subset of the
$M$-set $X$, we denote by $mK$ the set $\{mx\mid x\in K\}$.}
\begin{equation}
    m^\prime K_m\subset K_{m^\prime m}\label{m'Km}
\end{equation}
for all $m^\prime\in M$.\footnote{On the face of it, we could also
consider families of sets of the form ${\bf K}_I:=\{K_m\mid m\in
I\}$ for any ideal $I$ in $M$, since \eq{m'Km} still makes sense
in this case. However, we can reduce this to the case with $I:=M$
by choosing $K_m$ to be the empty set for all $m\not\in I$.} Then
if we define (cf.\ \eq{Def:xinJ})
\begin{equation}
    [x\in{\bf K}]_{BM}:=\{m\in M\mid mx\in K_m\} \label{Def:[xinbfK]}
\end{equation}
it is easy to check that the right hand side of \eq{Def:[xinbfK]}
is a left ideal in $M$. Thus another structure that can give a
source of generalised truth values is a family of subsets
$\{K_m\subset X,m\in M\}$ that satisfies \eq{m'Km}.\footnote{With
some effort it can be shown that families $\{K_m,m\in M\}$
satisfying \eq{m'Km} are in one-to-one correspondence with
equivariant maps $\lambda:X\times M\rightarrow LM$. Specifically,
given such a map $\lambda$ define $K^\lambda_m:=\{x\in X\mid
\lambda(x,m)=1\}$. Conversely, given a family ${\bf K}=\{K_m,m\in
M\}$ satisfying \eq{m'Km} define $\lambda^{{\bf
K}}(x,m):=\{m^\prime\in M\mid m^\prime x\in K_{m^\prime m}\}$. The
significance of this result is that equivariant maps
$\lambda:X\times M\rightarrow LM$ correspond to the points (in the
ordinary set-theoretic sense) of the power object $PX$ of the
object $X$ in $BM$ \cite{Goldblatt}. This is an important part of
the general theory of the topos $BM$ but it has been relegated to
a footnote since I am trying to minimise the amount of `heavy'
mathematics in the main text.}

In particular, if $K$ is {\em any\/} subset of $X$ (not
necessarily $M$-invariant) and if we define $K_m:=mK$, we see at
once that \eq{m'Km} is satisfied. In short, any subset $K\subset
X$ gives rise to a generalised truth value\footnote{One must be
careful not to confuse \eq{Def:[xinK]} with \eq{Def:xinJ}. If $K$
is an $M$-invariant subset of $X$, the definition in
\eq{Def:[xinK]} still makes sense, but this is generally not the
same as \eq{Def:xinJ} since there will typically be elements $m\in
M$ such that $mK$ is a proper subset of $K$. When $K$ is an
invariant subset we will use \eq{Def:xinJ} (rather than
\eq{Def:[xinK]}) since this corresponds to thinking of $K$ as a
sub-object of $X$ in $BM$.}
\begin{equation}
    [x\in K]_{BM}:=\{m\in M\mid mx\in mK\}.   \label{Def:[xinK]}
\end{equation}
It can readily be checked that the right hand side of
\eq{Def:[xinK]} is indeed a left ideal in the monoid $M$. This
example will play a central role in the applications to quantum
theory.

More generally, if $K_1$, $K_2$ are any pair of subsets of $X$ we
can define
\begin{equation}
    [K_1\subset K_2]_{BM}:=\{m\in M\mid mK_1\subset mK_2\}.
\end{equation}

A particular example of \eq{Def:[xinK]} is $K:=\{y\}$ for some
$y\in X$. In this special case, \eq{Def:[xinK]} can be written as
\begin{equation}
    [x=y]_{BM}:=\{m\in M\mid mx=my\}.     \label{Def:[x=y]}
\end{equation}
The right hand side of \eq{Def:[x=y]} is clearly a left ideal in
$M$: for if $m\in M$ is such that $mx=my$ then, trivially,
$nmx=nmy$ for all $n\in M$. Thus \eq{Def:[x=y]} is a measure in
the topos of $M$-sets of the extent to which the points $x,y$ in
$X$ are  `partially equal'. Indeed, $[x=y]_{BM}$ is larger the
`closer' $x$ and $y$ are to being equal, with $[x=y]_{BM}=M$ (the
identity of the Heyting algebra $LM$) if $x=y$.

\subsection{A monoid concept of `nearness to truth' in classical
physics.}\label{SubSec:MonoidClassPhys} An application of a topos
of type $BM$ arises in classical physics. Here we have a classical
state space $\cal S$ (a smooth manifold) in which each physical
quantity $A$ is represented by a smooth, real-valued function,
$\overline A$, on $\cal S$. Each state $s\in\cal S$ gives rise to
a simple valuation on propositions of the form
\begin{equation}
 V^s(A\in\Delta):=\left\{
                \begin{array}{ll}
                    1 & \mbox{if $\overline
A(s)\in\Delta$;} \\[2pt]
                    0 & \mbox{otherwise.}
                \end{array}\label{Def:Vs(AinDelta)}
                \right.
\end{equation}
In other words, the proposition ``$A\in\Delta$'' is true if the
state $s$ is such that $\overline A(s)$ belongs to $\Delta$;
otherwise it is false. Equivalently, ``$A\in\Delta$'' is true if
and only if $s\in \overline A^{-1}(\Delta):=\{s\in{\cal S}\mid
\overline A(s)\in\Delta\}$.

Such a simple `either-or' perspective seems natural in the context
of classical physics, and indeed one may wonder what else the
proposition ``$A\in\Delta$'' could mean other than the information
conveyed by \eq{Def:Vs(AinDelta)}. All this seems clear-cut---but
is it really so? For suppose $s$ is a state that does not belong
to $\overline A^{-1}(\Delta)$ but which, nevertheless, is `almost'
in this subset (so that $\overline A(s)$ `almost' belongs to
$\Delta$): is there not then some sense in which the proposition
``$A\in\Delta$'' is `almost true'? Contrariwise, suppose $s$ is
such that $\overline A(s)$ belongs to $\Delta$, but only just so
(i.e., $\overline A(s)$ is `close' to the edges of $\Delta$): then
is not ``$A\in\Delta$'' `almost false', or `only just true'? Such
grey-scale judgements are made frequently in daily life, but at
first sight there seems to be no role for them in the harsh,
black-and-white mathematics of classical physics.

From a mathematical perspective, the problem is how to judge the
nearness of any point $s$ in $\cal S$ to the subset $\overline
A^{-1}(\Delta)$ of $\cal S$. Of course, we could always put a
metric on $\cal S$, but there is in general no obvious or natural
way of choosing this (notwithstanding the fact that, in classical
physics, $\cal S$ is a symplectic manifold with a canonical
two-form).

However, a more appealing approach is based on the observation
that if the state $s$ is such that $\overline A(s)\in\Delta$ then,
of necessity, $f(\overline A(s))\in f(\Delta)$ for any smooth
function $f:\mathR\rightarrow\mathR$. This type of coarse-graining
was discussed in detail in \cite{IB99} in the context of assigning
truth values to propositions ``$A\in\Delta$'' when the state of
the system is a macrostate $M\subset {\cal S}$. In the present
case, we have $M=\{s\}$, and then the analysis in \cite{IB99}
results in the generalised valuation\footnote{The coarse-graining
of the original proposition ``$A\in\Delta$'' that is implicit in
\eq{Def:Vs(AinDelta)Gen} can be seen by noting that $f(\overline
A(s))\in f(\Delta)$ if and only if $\overline A(s)\in
f^{-1}(f(\Delta))$, and hence \eq{Def:Vs(AinDelta)Gen} assigns to
the proposition ``$A\in\Delta$'' all those weaker (coarse-grained)
propositions ``$A\in f^{-1}(f(\Delta))$'' which are `true' in the
normal sense of the word.}
\begin{equation}
        V^s(A\in\Delta):=\{f\in C^\infty(\mathR,\mathR)\mid
        f(\overline A(s))\in f(\Delta)\}  \label{Def:Vs(AinDelta)Gen}
\end{equation}
where $C^\infty(\mathR,\mathR)$ denotes the set of smooth (i.e.,
infinitely differentiable) functions $f:\mathR\rightarrow\mathR$.

In \cite{IB99},  the discussion of \eq{Def:Vs(AinDelta)Gen}
employed a topos of presheaves with truth values being sieves.
However, \eq{Def:Vs(AinDelta)Gen} can easily be reinterpreted in
terms of a topos $BM$. Specifically, we note that, since the
composition of a pair of smooth functions is itself smooth, the
set $C^\infty(\mathR,\mathR)$ can be given a monoid structure
whose combination law is defined as $f\star g\,(r):=f(g(r))$ for
all $r\in\mathR$. We then see at once that the right hand side of
\eq{Def:Vs(AinDelta)Gen} is actually a {\em left ideal\/} in this
monoid. Indeed, if $f\in C^\infty(\mathR,\mathR)$ is such that
$f(\overline A(s))\in f(\Delta)$ then, trivially, for all
$h:\mathR\rightarrow\mathR$ we have $h(f(\overline A(s)))\in
h(f(\Delta))$. Thus $f(\overline A(s))\in f(\Delta)$ implies that,
for all $h\in C^\infty(\mathR,\mathR)$, we have $h\star
f\,(\overline A(s))\in h\star f\,(\Delta)$, which means precisely
that the right hand side of \eq{Def:Vs(AinDelta)Gen} is a left
ideal in the monoid $C^\infty(\mathR,\mathR)$.

This remark suggests that the generalised valuation in
\eq{Def:Vs(AinDelta)Gen} could be understood in terms of the topos
of $C^\infty(\mathR,\mathR)$-sets. This is indeed the case: in
particular, we consider the obvious action of the monoid
$C^\infty(\mathR,\mathR)$ on the set $\mathR$, defined
by\footnote{This is a special case of a much wider class of
examples. Indeed, for any set $X$ there is a natural action of the
monoid ${\rm Map}(X,X)$ on $X$ given by (cf.\ \eq{Def:ellf(r)})
$\ell_f(x):=f(x)$ for all $f\in{\rm Map}(X,X)$ and $x\in X$. If
$X$ is a topological space, it is natural to restrict attention to
the sub-monoid $C(X,X)$ of continuous functions from $X$ to $X$.
If $X$ is a differentiable manifold, one would use the sub-monoid
$C^\infty(X,X)$ of smooth functions from $X$ to $X$. Note that
these subsets of ${\rm Map}(X,X)$ are indeed sub-{\em monoids\/}
since the composition of a pair of continuous (resp.\ smooth)
functions is itself continuous (resp.\ smooth). More generally, if
$X$ is an object in an arbitrary (small) category with a terminal
object $1$, one could use the monoid ${\rm Hom}(X,X)$ of arrows
whose domain and range is $X$, and with the obvious action on the
global elements $x:1\rightarrow X$ in which $f\in {\rm Hom}(X,X)$
transforms $x$ to $f\circ x$.}
\begin{equation}
    \ell_f(r):=f(r)     \label{Def:ellf(r)}
\end{equation}
for all $f\in C^\infty(\mathR,\mathR)$ and $r\in\mathR$. Now, for
each fixed state $s$ in $\cal S$, $\overline A(s)$ belongs to
$\mathR$, and hence, applying \eq{Def:[xinK]} with $X:=\mathR$,
$x:=\overline A(s)$, and $K:=\Delta\subset\mathR$, we see that
\begin{equation}
    [\overline A(s)\in\Delta]_{BC^\infty(\mathR,\mathR)}
    = \{f\in C^\infty(\mathR,\mathR)\mid
        f(\overline A(s))\in f(\Delta)\}.
\end{equation}
In other words, the generalised valuation in
\eq{Def:Vs(AinDelta)Gen} is just $[\overline
A(s)\in\Delta]_{BC^\infty(\mathR,\mathR)}$.

With an eye to the application to quantum theory to be discussed
in Sec.\ \ref{SubSec:MonoidQT}, we note that another monoid
interpretation of \eq{Def:Vs(AinDelta)Gen} can be obtained by
considering the action of the monoid $C^\infty(\mathR,\mathR)$ on
the set $C^\infty({\cal S},\mathR)$ whose elements (smooth,
real-valued functions $\overline A$, $\overline B$ on $\cal S$)
represent physical quantities in the system. Specifically, we
define
\begin{equation}
    \ell_f(\overline B):=f\circ \overline B\label{Def:gammafB}
\end{equation}
for all $f\in C^\infty(\mathR,\mathR)$ and $\overline B\in
C^\infty({\cal S},\mathR)$. We can also define an action of
$C^\infty(\mathR,\mathR)$ on the family, $P(\mathR)$, of subsets
of $\mathR$ by
\begin{equation}
    \ell_f(\Gamma):=f(\Gamma)     \label{Def:gammafDelta}
\end{equation}
for all $f\in C^\infty(\mathR,\mathR)$ and $\Gamma\subset \mathR$.
These operations combine to give an action of the monoid
$C^\infty(\mathR,\mathR)$ on $C^\infty({\cal S},\mathR)\times
P(\mathR)$ defined by
\begin{eqnarray}
    \ell_f: C^\infty({\cal S},\mathR)\times
    P(\mathR)&\rightarrow& C^\infty({\cal S},\mathR)\times P(\mathR)
                    \nonumber\\
            (\overline B,\Gamma)&\mapsto& (f\circ \overline B,f(\Gamma))
                        \label{Def:gammafBDelta}
\end{eqnarray}
If desired, this can also be viewed as defining an action of
$C^\infty(\mathR,\mathR)$ on the space of propositions of the type
``$B\in\Gamma$''. In other words, the proposition ``$B\in\Gamma$''
is mapped by $f$ to the proposition ``$f(B)\in f(\Gamma)$''.

We then define, for each state $s\in\cal S$, the set
\begin{equation}
    E^s:=\{(\overline B,\Gamma)\mid \overline B(s)\in\Gamma\}
            \label{Def:Es}
\end{equation}
and note that this subset of $C^\infty({\cal S},\mathR)\times
P(\mathR)$ is {\em invariant\/} under the action of the monoid
$C^\infty(\mathR,\mathR)$ (for, if $\overline B(s)\in\Gamma$ then
certainly $f(B(s))\in f(\Gamma)$ for all $f\in
C^\infty(\mathR,\mathR)$). As such, it is a sub-object of
$C^\infty({\cal S},\mathR)\times P(\mathR)$ in the topos
$BC^\infty(\mathR,\mathR)$, and hence there is an associated
characteristic arrow from $C^\infty({\cal S},\mathR)\times
P(\mathR)$ to the set $LC^\infty(\mathR,\mathR)$ of left ideals of
$C^\infty(\mathR,\mathR)$. According to the general result in
\eq{Def:xinJ}, this gives rise to the generalised truth value
\begin{eqnarray}
    [(\overline A,\Delta)\in E^s]_{BC^\infty(\mathR,\mathR)}
        &=&\{f\in C^\infty(\mathR,\mathR)\mid
        (f\circ\overline A,f(\Delta))\in E^s\}\nonumber\\[3pt]
        &=& \{f\in C^\infty(\mathR,\mathR)\mid f(\overline
        A(s))\in f(\Delta)\}\label{[ADeltainEs]}
\end{eqnarray}
which is precisely the right hand side of the generalised
valuation \eq{Def:Vs(AinDelta)Gen}.

\subsection{Using the monoid $M(\mathR,\mathR)$ in quantum theory}
\label{SubSec:MonoidQT} We can now discuss a monoid
reinterpretation of the generalised valuation
\eq{Def:Vs(AinDelta)}\footnote{Note that the right hand side of
\eq{Def:VpsiAinDelta2} is invariant under the scaling
$\ket\psi\mapsto \lambda\ket\psi$ for all non-zero complex numbers
$\lambda$. Hence \eq{Def:VpsiAinDelta2} defines a valuation on the
projective Hilbert space $P\cal H$ of all rays in $\cal H$, and we
could just as well denote the left hand side as
$V^{[\ket\psi]}(A\in\Delta)$ where $[\ket\psi]$ denotes the ray
that passes through the vector $\ket\psi$.}
\begin{equation}
 V^{\ket\psi}(A\in\Delta):=
 \{f:\mathR\rightarrow\mathR\mid \hat E[f(A)\in
            f(\Delta)]\ket\psi=\ket\psi\}\label{Def:VpsiAinDelta2}
\end{equation}
that was introduced in \cite{IB98} in the context of our topos
analysis of the Kochen-Specher theorem. In that
earlier\footnote{See \cite{Doring05} for a recent, and very
sophisticated, analysis of the Kochen-Specher theorem using the
mathematics of presheaves.} paper, the right hand side of
\eq{Def:VpsiAinDelta2} was interpreted as a sieve of arrows on the
object $\hat A$ in a category\footnote{In \cite{IB98} the category
${\cal A}({\cal H})$ was denoted $\cal O$.} ${\cal A}({\cal H})$
whose objects are bounded, self-adjoint operators, and whose
arrows $f_{{\cal A}({\cal H})}:\hat A\rightarrow \hat B$ are
defined to be all real functions $f:\mathR\rightarrow\mathR$ with
the property that $\hat B=f(\hat A)$.

The underlying mathematics is, again, presheaf theory, but in the
light of the discussion above, it is reasonable to enquire if
\eq{Def:VpsiAinDelta2} can be reinterpreted in a monoid language.
To this end, first recall that if $\hat A$ is any bounded,
self-adjoint operator then, for any bounded, measurable function
$f:\mathR\rightarrow \mathR$, the operator $f(\hat A)$ can be
defined using the spectral theorem for $\hat A$, and this operator
is also bounded and self-adjoint. We denote the set of all such
functions $f:\mathR\rightarrow\mathR$ by $M(\mathR,\mathR)$, and
note that this can be given a monoid structure by composition
since the composition of any pair of bounded and measurable
functions is itself bounded and measurable.

Then the critical observation is that the right hand side of
\eq{Def:VpsiAinDelta2} is actually a {\em left ideal\/} in this
monoid. The reason is analogous to that in Section
\ref{SubSec:MonoidClassPhys} in regard to the discussion following
\eq{Def:Vs(AinDelta)Gen}. Specifically, for any $h\in
M(\mathR,\mathR)$, we have\footnote{\label{footnoteBorel}Strictly
speaking, $\Gamma$ has to be a Borel subset of $\mathR$ in order
for the spectral projector $\hat E[B\in\Gamma]$ to exist. However,
this then raises the difficulty that if $\Gamma$ is Borel it is
not necessarily the case that $h(\Gamma)$ is Borel for arbitrary
$h\in M(\mathR,\mathR)$. This issue is resolved in \cite{IB99} but
we will not dwell on it here.}
\begin{equation}
    \hat E[B\in\Gamma]\preceq E[h(B)\in h(\Gamma)]\label{EADprecEhA}
\end{equation}
in the partial ordering of the lattice of projection operators. It
follows at once that if $\ket\psi$ and $f$ are such that $\hat
E[f(A)\in f(\Delta)]\ket\psi=\ket\psi$ then $\hat E[h(f(A))\in
h(f(\Delta))]\ket\psi=\ket\psi$ for all $h\in M(\mathR,\mathR)$.
But this is precisely the statement that \eq{Def:VpsiAinDelta2} is
a left ideal in the monoid $M(\mathR,\mathR)$.

This suggests strongly that the generalised valuation
\eq{Def:VpsiAinDelta2} can be reinterpreted using the language of
the topos of $M(\mathR,\mathR)$-sets. To complete this
identification it is necessary to find an appropriate set on which
the monoid $M(\mathR,\mathR)$ acts, and then apply the general
result in \eq{Def:xinJ}.

 The first relevant observation is that if ${\cal
A}({\cal H})$ denotes the set of all bounded, self-adjoint
operators on $\cal H$, then the operation whereby $\hat B\in {\cal
A}({\cal H})$ is replaced by $f(\hat B)$, with $f\in
M(\mathR,\mathR)$, can be viewed as a left action of the monoid
$M(\mathR,\mathR)$ on ${\cal A}({\cal H})$ (cf.\
\eq{Def:gammafB}). Similarly, if $B(\mathR)$ denotes the
collection of bounded, Borel subsets of $\mathR$, then an action
of $M(\mathR,\mathR)$ on $B(\mathR)$ can be defined\footnote{It is
necessary to take into account the cautionary remark in footnote
\ref{footnoteBorel}.} by letting $f\in M(\mathR,\mathR)$ take
$\Gamma\in B(\mathR)$ to $f(\Gamma)$. This is a direct analogue of
the action, \eq{Def:gammafDelta}, in the classical case.

Combining these two operations gives a left action of the monoid
$M(\mathR,\mathR)$ on ${\cal A}({\cal H})\times B(\mathR)$ which
is defined as (cf.\ \eq{Def:gammafBDelta})
\begin{eqnarray}
    \ell_f:{\cal A}({\cal H})\times B(\mathR)&\rightarrow&
                    {\cal A}({\cal H})\times B(\mathR)\nonumber\\
        (\hat B,\Gamma)&\mapsto& (f(\hat B), f(\Gamma))
\end{eqnarray}
for all $f\in M(\mathR,\mathR)$. This can also be viewed as an
action of $M(\mathR,\mathR)$ on the space of propositions of the
form ``$B\in\Gamma$''. In any event, what is important is that to
each vector $\ket\psi\in\cal H$, we can define (cf.\ \eq{Def:Es})
\begin{equation}
    E^{\ket\psi}:=\{(\hat B,\Gamma)\mid \hat E[B\in\Gamma]\ket\psi
    =\ket\psi\}. \label{Def:Epsi}
\end{equation}
Then the crucial observation is that this subset of ${\cal
A}({\cal H})\times B(\mathR)$ is {\em invariant\/} under the
action of the monoid $M(\mathR,\mathR)$. This follows at once from
the partial ordering relation in \eq{EADprecEhA} which guarantees
that if $(\hat B,\Gamma)\in E^{\ket\psi}$ (so that $\hat
E[B\in\Gamma]\ket\psi =\ket\psi)$ then $(f(\hat B),f(\Gamma))\in
E^{\ket\psi}$ for all $f\in M(\mathR,\mathR)$.

We can now use the general definition in \eq{Def:xinJ} to compute
this subset's characteristic function from ${\cal A}({\cal
H})\times B(\mathR)$ to $LM(\mathR,\mathR)$. This gives
\begin{eqnarray}
    [(\hat A,\Delta)\in E^{\ket\psi}]_{BM(\mathR,\mathR)}
        &:=&\{f\in M(\mathR,\mathR)\mid (f(\hat
    A),f(\Delta))\in E^{\ket\psi}\}\\
                    &=& \{f\in M(\mathR,\mathR)\mid
    \hat E[f(A)\in f(\Delta)]\ket\psi=\ket\psi\}
\end{eqnarray}
which is precisely the right hand side of \eq{Def:VpsiAinDelta2}.
Thus the generalised truth value in \eq{Def:VpsiAinDelta2} has an
interpretation in terms of the topos of $M(\mathR,\mathR)$-sets.
Note the close analogy with the result \eq{[ADeltainEs]} of the
classical theory.

\section{A topos interpretation of state-vector reduction}
\label{Sec:ToposReduction}
\subsection{Actions of the monoid $\LH$}
So far, our application of monoid theory to quantum mechanics has
been to use the language of $M(\mathR,\mathR)$-sets to re-express
earlier results obtained originally using presheaf theory.
However, we wish now to develop a new, `neo-realist'
interpretation of quantum theory that uses a topos of $M$-sets in
a fundamental way.

Given a quantum theory with a Hilbert space $\cal H$, one obvious
monoid to consider is the set $\LH$ of all bounded, linear
operators on $\cal H$. The monoid composition law is the operator
product, and the unit element is simply the unit operator $\hat
1$. A related monoid is obtained by defining two operators to be
equivalent, $\hat A\equiv\hat B$, if there exists
$\lambda\in\mathC_*$ (the non-zero complex numbers) such that
$\hat A=\lambda\hat B$. We denote the set of equivalence classes
as $\LH/\mathC_*$ and note that this can be given a monoid
structure with the combination law
\begin{equation}
    [\hat A][\hat B]:=[\hat A\hat B]
\end{equation}
where $[\hat A]$ denotes the equivalence class of $\hat A$. This
particular monoid was discussed in quantum theory many years ago
in the context of the theory of Baer $*$-semigroups \cite{Pool75}
\cite{BC81}.

The obvious set on which the monoid $\LH$ acts is $\cal H$ itself,
with $\ell_{\hat A}(\ket\psi):=\hat A\ket\psi$ for all $\hat A\in
\LH$ and $\ket\psi\in\cal H$.

Another natural action is on the projective Hilbert space $P{\cal
H}$, with the action on any ray $[\ket\psi]$ (the ray that passes
through the (non-null) vector $\ket\psi$) being
\begin{equation}
\ell_{\hat A}([\ket\psi]):=[\hat A\ket\psi] \label{Def:ellApsi}
\end{equation}
Here, the meaning of the symbol $[\hat A \ket\psi]$ is as follows.
If $\hat A\ket\psi\neq 0$, then $[\hat A\ket\psi]$ denotes the ray
that passes through $\hat A\ket\psi$. However, if $\hat
A\ket\psi=0$, then $[\hat A\ket\psi] =[0]$ denotes a special point
that must be added to the projective Hilbert space. Thus the
action of our monoid is not on $P{\cal H}$ but on $P{\cal
H}\cup[0]$. Of course, $\ell_{\hat A}[0]=[0]$ for all operators
$\hat A$ in the monoid $\LH$. In other words, $[0]$ is an {\em
absorbing\/} element for the action of $\LH$ on $P{\cal
H}\cup[0]$.

The action on vectors can be extended to give an action of the
monoid $\LH$ on arbitrary closed, linear subspaces of $\cal H$.
Specifically, if $K\subset\cal H$ is such a subspace then, for all
$\hat A\in\LH$, we define
\begin{equation}
    \ell_{\hat A}(K):=\hat AK:=\{\hat A\ket\psi\mid \ket\psi\in
    K\}^{\rm cl}     \label{Def:gammaAK}
\end{equation}
where the superscript $\{\,\}^{\rm cl}$ signifies that the
topological closure is to be taken of the quantity inside the
parentheses. Note that since there is a one-to-one correspondence
between closed, linear subspaces on $\cal H$ and projection
operators, \eq{Def:gammaAK} also generates an action of the monoid
$\LH$ on the collection of projectors. However, there is no
obvious way of writing down explicitly what $\hat A$ does to any
particular projector.

From a projective perspective, we denote by $PK$ the set of all
rays passing through the non-null vectors in $K$. We then get an
action of $L({\cal H})$ on $PK\cup[0]$ defined by
\begin{equation}
        \ell_{\hat A}(PK):=\bigcup_{[\ket\psi]\in PK}{[\hat A\ket\psi]}
\end{equation}
and with $\ell_{\hat A}[0]:=0$ as before.

If one thinks of quantum states as being represented by normalised
vectors then one might try to define an action of $\LH$ by
\begin{equation}
    \ket\psi\mapsto {{\hat A\ket\psi}\over\|\hat A\ket\psi\|}.
                \label{Def:gammaApsi}
\end{equation}
Note that the right hand side of \eq{Def:gammaApsi} is invariant
under the transformation $\hat A\mapsto \lambda\hat A$,
$\lambda\in\mathC_*$. Thus \eq{Def:gammaApsi} passes to an action
of the monoid $L({\cal H})/\mathC_*$. There is an analogue of
\eq{Def:gammaApsi} on density matrices
\begin{equation}
    \hat\rho\mapsto {\hat A\rho\hat A^\dagger\over {\rm tr}(\hat\rho
    \hat A\hat A^\dagger)}.\label{Def:gammaArho}
\end{equation}

We note however that \eq{Def:gammaApsi} is only defined if $\hat
A\ket\psi\neq 0$, and similarly \eq{Def:gammaArho} requires ${\rm
tr}(\hat\rho\hat A\hat A^\dagger)\neq 0$. This means that neither
\eq{Def:gammaApsi} or \eq{Def:gammaArho} corresponds to a
well-defined action of the monoid $\LH$: we shall return to this
problem later.

\subsection{Truth values using the monoid $L({\cal H})$}
Let us now consider how truth values in the set of left ideals of
$L({\cal H})$ could arise. One of the simplest expressions is
\eq{Def:[x=y]} which, for the monoid action of $\LH$ on $\cal H$,
reads
\begin{eqnarray}
    \Big[\ket\psi=\ket\phi\Big]_{B\LH}&:=&\{\hat B\in\LH\mid \hat B\ket\psi=\hat
    B\ket\phi\} \\ \label{Def:[psi=phi]}
        &=&\{\hat B\in\LH\mid \hat B(\ket\psi-\ket\phi)=0\}
\end{eqnarray}
which is clearly a left ideal of $\LH$.  There is an analogous
expression on the extended projective Hilbert space (i.e., on
$P{\cal H}\cup[0]$) of the form
\begin{equation}
    \Big[[\ket\psi]=[\ket\phi]\Big]_{B\LH}:=\{\hat B\in\LH\mid [\hat
    B\ket\psi]=[\hat B\ket\phi]\,\}.
\end{equation}
Note that the equation $[\hat B\ket\psi]=[\hat B\ket\phi]$ implies
that $\hat B\ket\psi=0$ if, and only if, $\hat B\ket\phi=0$.

From a mathematical perspective, \eq{Def:[psi=phi]} is an
interesting Heyting-algebra valued measure of the extent to which
the vectors $\ket\psi$ and $\ket\phi$ are not equal. However, as
it stands, it is hard to give any physical meaning to this
expression. Basically, the problem is that the monoid $\LH$
consists of {\em all\/} bounded operators, whereas, in quantum
theory, the most important operators are unitary operators and
self-adjoint operators.

We could consider the sub-monoid of unitary operators, but this is
uninteresting since a unitary operator is invertible, and hence
one-to-one. This means that, for example, the analogue of
\eq{Def:[psi=phi]} for unitary operators is the empty set unless
$\ket\psi=\ket\phi$.

One might be tempted to consider the collection ${\cal A}({\cal
H})$ of bounded, self-adjoint operators on $\cal H$, but this is
not a sub-monoid of $\LH$ since the product of self-adjoint
operators is not itself self-adjoint unless they commute. However,
this remark suggests another possibility which, it transpires,
{\em is\/} fruitful: namely, consider the subset, ${\rm Pr}{\cal
A}({\cal H})$, of $\LH$ consisting of all finite products of
self-adjoint operators. This {\em is\/} a sub-monoid, and gives
rise to the expression
\begin{equation}
    \Big[\ket\psi=\ket\phi\Big]_{B{\rm Pr}{\cal A}({\cal H})}:=
    \{\hat A_n\hat A_{n-1}\cdots\hat A_1\mid
            \hat A_n\hat A_{n-1}\cdots\hat A_1\ket\psi=
   \hat A_n\hat A_{n-1}\cdots\hat A_1\ket\phi\}.\label{[psi=phi]PrA}
\end{equation}

This expression still has no obvious physical meaning, but it does
suggest one thing very strongly: namely, the process of
state-vector reduction! This is the procedure whereby if a series
of (ideal) measurements is made of physical quantities whose
corresponding outcomes are represented by the projection operators
$\hat P_1,\hat P_2,\ldots,\hat P_n$ respectively, then after the
measurements are made (neglecting time development between them)
the state vector has been reduced to
\begin{equation}
    \ket\psi \mapsto \hat P_n\hat P_{n-1}\cdots \hat P_1\ket\psi.
                \label{RedKetpsi}
\end{equation}
Of course, this can be viewed as the result of a series of
reductions
\begin{equation}
    \ket\psi\mapsto \hat P_1\ket\psi\mapsto \hat P_2\hat
    P_1\ket\psi \mapsto\cdots \mapsto \hat P_n\hat P_{n-1}\cdots \hat
    P_1\ket\psi.\label{StringReductions}
\end{equation}

Actually, \eq{RedKetpsi} is not quite correct, as we need to
consider the normalisation of the reduced vector. For the moment
though, we can say that the key idea is to think of the reduction
\eq{RedKetpsi} as being the result of an action on $\cal H$ of the
sub-monoid\footnote{The notation is potentially confusing here.
The symbol $P{\cal H}$ denotes the projective Hilbert space---{\em
i.e.}, the space of (complex) one-dimensional subspaces of $\cal
H$; on the other hand, $P({\cal H})$ denotes the space of
projection operators on $\cal H$.}, ${\rm Pr}P({\cal H})$, of
finite products of projection operators.

For this particular monoid, the general equation \eq{Def:[x=y]}
reads
\begin{equation}
\Big[\ket\psi=\ket\phi\Big]_{B{\rm Pr}P({\cal H})}:=\{\hat P_n\hat
P_{n-1}\cdots\hat P_1\mid
            \hat P_n\hat P_{n-1}\cdots\hat P_1\ket\psi=
            \hat P_n\hat P_{n-1}\cdots\hat P_1\ket\phi\}
                    \label{[psi=phi]PrP}
\end{equation}
or, perhaps better, we should use rays in the Hilbert space and
define
\begin{equation}
\Big[\,[\!\ket\psi]=[\!\ket\phi]\,\Big]_{B{\rm Pr}P({\cal
H})}:=\{\hat P_n\hat P_{n-1}\cdots\hat P_1\mid
            [\hat P_n\hat P_{n-1}\cdots\hat P_1\ket\psi]=
            [\hat P_n\hat P_{n-1}\cdots\hat P_1\ket\phi]\,\}
                    \label{[psi=phi]PrPRay}
\end{equation}
Note that the right hand side of \eq{[psi=phi]PrPRay} is
equivalent to the statement that there exists $\lambda\in\mathC_*$
such that $\hat P_n\hat P_{n-1}\cdots\hat P_1\ket\psi =\lambda
\hat P_n\hat P_{n-1}\cdots\hat P_1\ket\phi$.

Unlike \eq{[psi=phi]PrA}, the expressions in \eq{[psi=phi]PrP} and
\eq{[psi=phi]PrPRay} {\em do\/} have a very interesting physical
interpretation. Namely, they assign (in a slightly different way)
as a measure of the similarity between two state vectors the
collection of those series of ideal measurements which, {\em if\/}
they were performed, give reduced vectors that can no longer be
distinguished from each other.

Strictly speaking, this is not quite correct, and will be amended
shortly in Section \ref{SubSec:MonoidStringsProjs}. However,
before doing that we note that this idea can be developed
immediately to attain our goal of producing a new type of truth
value for propositions ``$A\in\Delta$'' in quantum theory. For let
${\cal H}_{A\in\Delta}$ denote the subspace of $\cal H$ that is
the image of the spectral projector $\hat E[A\in\Delta]$;  i.e.,
the proposition ``$A\in\Delta$'' is true with probability $1$ for
all states $\ket\phi$ in ${\cal H}_{A\in\Delta}$. Then, based on
the general result \eq{Def:[xinK]}, we can define the new
generalised valuation
\begin{eqnarray}
\lefteqn{V^{\ket\psi}(A\in\Delta)_{B{\rm Pr}P({\cal
H})}:=[\ket\psi\in{\cal
H}_{A\in\Delta}]_{B{\rm Pr}P({\cal H})}}\ \nonumber\\
     &&= \{\hat P_n\hat P_{n-1} \cdots\hat P_1\mid
            \hat P_n\hat P_{n-1}\cdots\hat P_1\ket\psi\in
            \hat P_n\hat P_{n-1}\cdots\hat P_1{\cal
            H}_{A\in\Delta}\}.\ \label{MVpsiAinDelta}
\end{eqnarray}

Alternatively, and probably better in terms of physical meaning,
we can adopt the projective perspective and define
\begin{eqnarray}
\lefteqn{V^{[\ket\psi]}(A\in\Delta)_{B{\rm Pr}P({\cal
H})}:=\Big[\,[\!\ket\psi]\in P{\cal
H}_{A\in\Delta}\Big]_{B{\rm Pr}P({\cal H})}}\ \nonumber\\
     &&= \{\hat P_n\hat P_{n-1} \cdots\hat P_1\mid
            [\hat P_n\hat P_{n-1}\cdots\hat P_1\ket\psi]\in
            \ell_{\hat P_n\hat P_{n-1}\cdots\hat P_1}(P{\cal
            H}_{A\in\Delta})\}.\ \label{MVpsiAinDeltaProj}
\end{eqnarray}
Note that if ${\cal H}_{A\in\Delta}$ is a one-dimensional subspace
of $\cal H$, then \eq{MVpsiAinDeltaProj} is equivalent to
\eq{[psi=phi]PrPRay}.

It must be emphasised that the assignment in \eq{MVpsiAinDelta} is
intended to be {\em counterfactual\/}: we are not interested in
state-vector reduction as it is normally understood, whether---as
in the instrumentalist interpretation of quantum theory---it is
regarded as a result of sub-ensemble selection, or whether---as in
more adventurous interpretations---it is interpreted either as an
effective physical process brought about by, for example,
decoherence, or as an actual physical process associated with some
non-linear modification of the Schr\"odinger equation. Rather, the
intention is to assign the left ideal \eq{MVpsiAinDelta} (and
similarly for \eq{MVpsiAinDeltaProj}) in the monoid ${\rm
Pr}P({\cal H})$ as the truth value of the proposition
``$A\in\Delta$'' in the state $\ket\psi$ with the intent of
producing a new type of `neo-realist' interpretation of the
quantum formalism: i.e., it is a non-standard (in the logical
sense) way of saying `how things are' in regard to the quantity
$A$ when the state is $\ket\psi$.

\subsection{The monoid of strings of projectors}
\label{SubSec:MonoidStringsProjs} At this point we should address
a small defect in the formalism as presented so far. Namely, given
a product $\hat P_n\hat P_{n-1}\cdots\hat P_1$ of projectors, it
is not possible to recover the individual projectors from this
operator since many different collections of projectors have the
same product. In this sense, the statement above that
\eq{[psi=phi]PrP} ``assigns as a measure of the similarity between
two state vectors the collections of those series of ideal
measurements\ldots'' is not strictly correct, and the formalism
must be modified slightly to gain the desired counterfactual
interpretation of \eq{[psi=phi]PrP} and \eq{MVpsiAinDelta} (or
\eq{MVpsiAinDeltaProj}) in terms of strings of possible
operations. This is done as follows.

The key idea is to construct a new monoid, $\SP$, whose elements
are finite strings of (non zero) projection operators, $R:=(\hat
R_p,\hat R_{p-1},\ldots, \hat R_1)$ ($p$ is called the {\em
length\/} of the string) and with the monoid product law defined
by concatenation of the strings. Thus if $R:=(\hat R_p,\hat
R_{p-1},\ldots,\hat R_1)$ and $Q:=(\hat Q_q,\hat
Q_{q-1},\ldots,\hat Q_1)$ we define the product as
\begin{equation}
    Q\star R:=(\hat Q_q,\hat Q_{q-1},\ldots,\hat Q_1,
            \hat R_p,\hat R_{p-1},\ldots,\hat R_1). \label{Def:Q*R}
\end{equation}
The unit element in the monoid $\SP$ is the empty string,
$\emptyset$. Physically, we think of the string $(\hat R_p,\hat
R_{p-1},\ldots, \hat R_1)$ as referring (counterfactually) to a
situation in which the first operation corresponds to the
projector $\hat R_1$, the second operation to $\hat R_2$, and so
on.

If $R:=(\hat R_p,\hat R_{p-1},\ldots,\hat R_1)$ belongs to $\SP$,
we define the {\em reduction\/} of $R$ to be the operator
\begin{equation}
    \hat R:= \hat R_p\hat R_{p-1}\cdots \hat R_1.
\end{equation}
As a matter of convention, we define $\hat\emptyset:=\hat 1$, so
that the unit element in the monoid $\SP$ reduces to the unit
operator. Note that $\widehat {Q\star R}=\hat Q\hat
R$.\footnote{Note also that we allow consecutive repetition of
projections operators in a string although, of course, the
reduction of a string with a repeated projector is the same as
that without.}

We can now return to our ideas about generalised valuations in
quantum theory and start by allowing the monoid $\SP$ to act on
$\cal H$ by
\begin{equation}
    \ell_Q(\ket\psi):=\hat Q\ket\psi
\end{equation}
for all finite strings $Q$ of projectors. The expression
\eq{[psi=phi]PrPRay} then gets replaced by
\begin{equation}
\Big[\,[\!\ket\psi]=[\!\ket\phi]\,\Big]_{B\SP}:=
        \{Q\in\SP\mid [\hat Q\ket\psi]= [\hat Q\ket\phi]\,\}\
\end{equation}
and the valuation in \eq{MVpsiAinDeltaProj} becomes
\begin{equation}
    V^{[\ket\psi]}(A\in\Delta)_{B\SP}:=
\{Q\in\SP\mid [\hat Q\ket\psi]\in
    \ell_{\hat Q}(P{\cal H}_{A\in\Delta})\}\
        \label{Def:VpsiAinDeltaSP}
\end{equation}
As desired, this is a left ideal in the monoid $\SP$, and thereby
gives a new generalised truth value for the proposition
``$A\in\Delta$'' in the quantum state $\ket\psi$.

\subsection{The question of normalisation}
If we think of a state of a quantum system as being determined by
a normalised vector $\ket\psi$, then strictly speaking the state
vector reduction should not be \eq{RedKetpsi} but rather
\begin{equation}
\ket\psi \mapsto {\hat P_n\hat P_{n-1}\cdots \hat P_1\ket\psi\over
        \|\hat P_n\hat P_{n-1}\cdots \hat P_1\ket\psi\|}
                \label{RedKetpsiNorm}
\end{equation}
which is fine as long as $\|\hat P_n\hat P_{n-1}\cdots \hat
P_1\ket\psi\|\neq 0$.  This is no problem in the conventional
formalism since, there, one never gets reduction to an eigenstate
for which there is {\em zero\/} probability of finding the
associated eigenvalue. Or, more precisely: such zero probability
events are swept under the carpet as never happening. However, for
our neo-realist view, the normalisation problem is a genuine issue
since in the action of the monoid $\SP$ on a state $\ket\psi$,
there will of course be strings $Q$ for which $\hat Q\ket\psi=0$.

There is an analogous normalisation issue for density matrices. In
order to extend the formalism to include density-matrix states, we
note first that the condition on the right hand side of the
non-projective version of \eq{Def:VpsiAinDeltaSP} would be $\hat
Q\ket\psi\in\hat Q{\cal H}_{A\in\Delta}$, and this is equivalent
to the statement that
\begin{equation}
    \ell_Q(\hat E[A\in\Delta])\hat Q\ket\psi=\hat Q\ket\psi
                \label{gammaQEAinDelta}
\end{equation}
where, in accordance with the remark following \eq{Def:gammaAK},
$\ell_Q(\hat E[A\in\Delta])$ denotes the projection operator onto
the subspace $\hat Q{\cal H}_{A\in\Delta}$. In turn,
\eq{gammaQEAinDelta} is equivalent to\footnote{If $\hat P$ is any
projector, and $\ket\phi$ is any vector, it follows from the
Schwarz inequality that $\hat P\ket\phi=\ket\phi$ is equivalent to
$\bra\phi\hat P\ket\phi=\bra\phi\!\phi\rangle$.}
\begin{equation}
    \bra\psi\hat Q^\dagger \ell_Q(\hat E[A\in\Delta])\hat
    Q\ket\psi =\bra\psi \hat Q^\dagger \hat Q\ket\psi.
        \label{brapsigammaketapsi}
\end{equation}

Rewriting \eq{gammaQEAinDelta} in the form of
\eq{brapsigammaketapsi} suggests how to extend the formalism to
include states that are density matrices. We can define an action
of the monoid $\SP$ on the set of hermitian, trace-class operators
$\hat \rho$ (i.e., the trace of $\rho$ exists as a finite real
number) by
\begin{equation}
    \ell_Q(\hat\rho):=\hat Q\hat\rho \hat Q^\dagger.
        \label{Def:gammaQrho}
\end{equation}
Of course, if $\hat\rho$ is a density matrix state (so that ${\rm
tr}(\hat\rho)=1$) then we might want to define a normalised
version of \eq{Def:gammaQrho} as
\begin{equation}
\ell_Q(\hat\rho):={\hat Q\hat\rho \hat Q^\dagger\over
                {\rm tr}(\hat Q\hat\rho \hat Q^\dagger)}
        \label{Def:gammaQrhoNorm}
\end{equation}
but this only makes sense if ${\rm tr}(\hat Q\hat\rho \hat
Q^\dagger)\neq 0$. However, we can avoid this difficulty by
imitating \eq{brapsigammaketapsi} and defining the generalised
valuation
\begin{equation}
    V^{\hat\rho}(A\in\Delta)_{\SP}:=\{Q\in \SP\mid
    {\rm tr}(\hat Q\hat\rho\hat Q^\dagger \ell_Q(\hat
    E[A\in\Delta]))={\rm tr}(\hat Q\hat\rho \hat Q^\dagger)\}.
    \label{Def:VrhoAinDeltaSP}
\end{equation}

\subsection{A new category to handle the normalisation issue}
\label{SubSec:NewCat} The trick used above to avoid the
normalisation issue does not negate the fact that the right hand
side of \eq{Def:VpsiAinDeltaSP} (resp.\ \eq{Def:VrhoAinDeltaSP})
necessarily includes strings $Q$ for which $\hat Q\ket\psi=0$
(resp.\ $\hat Q\hat\rho\hat Q^\dagger=0$). Whether or not this is
problematic is somewhat debatable. On the one hand, it is true
that, as has been remarked already, in the conventional formalism
such zeros do not occur. On the other hand, our monoid methods are
aimed at giving a neo-realist interpretation of quantum theory,
and, as such, it is not {\em a priori\/} necessary that they
replicate exactly the structure of state-vector reduction in the
conventional formalism. In that sense, the mathematics, as it is,
does work.

However, if the normalisation question {\em is\/} thought to be a
genuine issue, then the first step might well seem to be that we
should restrict our attention to strings $Q:=(\hat Q_q,\hat
Q_{q-1},\ldots, \hat Q_1)$ for which $\hat Q:=\hat Q_q\hat
Q_{q-1}\cdots \hat Q_1 \neq 0$. We shall denote the set of all
such strings by $\SP_0$. Thus the elements of $\SP_0$ have the
property that they do not cause difficulties for {\em any\/}
vectors in $\cal H$.

The problem, however, is that if $Q_1$ and $Q_2$ are members of
$\SP_0$, their monoid product $Q_2Q_1$ may not have this property.
For example, considered as strings of unit length, any non-null
projectors $\hat P$, $\hat Q$ belong to $\SP_0$, but if $\hat P$
and $\hat Q$ are orthogonal then $\hat Q\hat P=0$.

This means that $\SP_0$ is only a {\em partial\/} monoid, with the
product $Q_2Q_1$ being defined only if $\hat Q_2\hat Q_1\neq 0$.
There are several ways in which this problem can be tackled, and I
will outline two of them here. A key observation is that a natural
source of partial monoids is category theory, since the arrows in
any category form a partial monoid: the composition $g\circ f$ of
any two arrows $f,g$ is only defined if the range of $f$ is equal
to the domain of $g$.  This suggests trying to associate the
elements of $\SP_0$ with the arrows in some category. One way is
to define a new category $\cal X$ as follows:
\begin{enumerate}
    \item[(i)] The objects are collections, $\Xi$, of non-zero vectors in
$\cal H$ with the property that if $\ket\psi\in\Xi$, then,
$\lambda\ket\psi\in\Xi$ for all $\lambda\in\mathC_*$.\footnote{It
also would be possible to consider the objects to be subsets of
rays. The analogue of our discussion for that case is obvious.}
    \item[(ii)] If $\Xi_1$ and $\Xi_2$ are a pair of objects, we
define the arrows between them as the elements of the set
\begin{eqnarray}
    {\rm Hom}(\Xi_1,\Xi_2)&:=&\{ Q\in\SP_0\mid\forall \ket\psi\in\Xi_1,
           \;  \hat Q\ket\psi\in\Xi_2\}\\
        &\equiv&\{Q\in\SP_0\mid \hat Q\Xi_1\subset\Xi_2\}.
\end{eqnarray}
If $Q\in {\rm Hom}(\Xi_1,\Xi_2)$ and $R\in {\rm Hom}(\Xi_2,\Xi_3)$
then the composite arrow $R\circ Q\in {\rm Hom}(\Xi_1,\Xi_3)$ is
simply the concatenation of the strings.
\end{enumerate}

Now, if $\ket\psi$, $\ket\phi$ belong to some object $\Xi$ we can
define, provisionally,
\begin{equation}
        [\ket\psi=\ket\phi]_{{\cal X},\Xi}:=
            \{Q\in{\rm Hom}(\Xi,\cdot)\mid \hat Q\ket\psi =
            \hat Q\ket\phi\}\label{Def:[psi=phi]XQProv}
\end{equation}
where ${\rm Hom}(\Xi,\cdot)$ denotes the set of all arrows whose
domain is $\Xi$. However, since the states concerned all have
non-zero norm, it is better to replace \eq{Def:[psi=phi]XQProv}
with the normalised form (and referring now to rays in the Hilbert
space)
\begin{equation}
        \Big[[\ket\psi]=[\ket\phi]\Big]_{{\cal X},\Xi}:=
            \left\{Q\in{\rm Hom}(\Xi,\cdot)\mid\exists z\in\mathC,
|z|=1, {\hat Q\ket\psi\over\|\hat Q\ket\psi\|} = z{\hat
Q\ket\phi\over\|\hat Q\ket\phi\|}\right\}\label{Def:[psi=phi]XQ}
\end{equation}
which, of course, is not equivalent to \eq{Def:[psi=phi]XQProv}
(we include the $z$ phase factor since normalised states are only
determined up to such factors). In fact, the condition on the
right hand side of \eq{Def:[psi=phi]XQ} is equivalent to the
statement that there exists some $\lambda\in\mathC_*$ such that
$\hat Q\ket\psi=\lambda\hat Q\ket\phi$.

It is clear that the right hand side of \eq{Def:[psi=phi]XQ} is a
sieve of arrows on $\Xi$, and hence a member of the Heyting
algebra of all sieves on $\Xi$; as such it is a possible
generalised truth value. It is easy to see how this would be
extended to give generalised truth values to propositions
``$\ket\psi\in K$'' for a linear subspace $K\subset\cal H$; in
particular to subspaces of $\Xi$ of the form ${\cal
H}_{A\in\Delta}$. Namely, as:
\begin{equation}
    V^{\ket\psi}(A\in\Delta)_{{\cal X},\Xi}:=
        \{Q\in{\rm Hom}(\Xi,\cdot)\mid \hat Q\ket\psi\in
            \hat Q{\cal H}_{A\in\Delta}\}.
                \label{Def:[VpsiAinDeltaXQ}
\end{equation}
Note that if ${\cal H}_{A\in\Delta}$ is a one-dimensional subspace
of $\cal H$, then \eq{Def:[VpsiAinDeltaXQ} is equivalent to
\eq{Def:[psi=phi]XQ}. Note also that \eq{Def:[psi=phi]XQ} and
\eq{Def:[VpsiAinDeltaXQ} are `contextual' in the sense that their
right hand sides depend on the subset $\Xi$ of vectors that is
chosen to contain $\ket\psi$, as well as ${\cal H}_{A\in\Delta}$
of course.

To give more meaning to this construction we observe that there is
an implicit `polar operation' at play here, as encapsulated in the
definition
\begin{equation}
    \Xi^0:=\{Q\in\SP_0\mid \forall\ket\psi\in\Xi,\; \hat Q\ket\psi
        \neq 0\}.    \label{Def:Xi0}
\end{equation}
Note that ${\rm Hom}(\Xi,\cdot)=\Xi^0$.

Similarly, if $J$ is a subset of $\SP_0$, we can define
\begin{equation}
    J^0:=\{\ket\psi\in{\cal H}_*\mid \forall Q\in J,\;
        \hat Q\ket\psi\neq 0\}\label{Def:J0}
\end{equation}
where ${\cal H}_*$ denotes the set of all non-null vectors in
$\cal H$. We note that $J_1\subset J_2$ implies $J_2^0\subset
J_1^0$; similarly $\Xi_1\subset\Xi_2$ implies
$\Xi_2^0\subset\Xi_1^0$. This is one reason for referring to these
operations as `polar'. Another is the fact that
$\Xi_1^0\cap\Xi_2^0=(\Xi_1\cup\Xi_2)^0$ for all objects $\Xi_1$
and $\Xi_2$, and similarly for pairs $J_1$ and $J_2$. We note that
this construction can also be understood in the language of {\em
Galois connections\/} \cite{Bell88} (which, in turn, are a special
case of adjoint functors) defined on the partially ordered sets
given by the subsets of $\SP_0$ and the subsets of ${\cal
H}_*$.\footnote{I thank Jeremy Butterfield for bringing this to my
attention. For an application of the theory of Galois connections
in standard quantum logic see \cite{Butterfield93}.}

We next note that
\begin{eqnarray}
    (\Xi^0)^0&=&\{\ket\psi\in{\cal H}_*\mid\forall Q\in\Xi^0,\;
                            \hat Q\ket\psi\neq 0\}\nonumber\\
            &=&\{\ket\psi\in{\cal H}_*\mid \forall\ket\phi\in\Xi,
                  \;  \hat Q\ket\phi\neq 0\Rightarrow\hat
Q\ket\psi\neq 0\}.\label{Xi00}
\end{eqnarray}
In particular, $\Xi\subset(\Xi^0)^0$. In fact, $(\Xi^0)^0$ is a
natural extension\footnote{In the language of Galois connections,
$(\Xi^0)^0$ is the `closure' of $\Xi$.} of the subset of vectors
$\Xi$ in the sense that we can extend $\Xi\subset{\cal H}_*$ to
$(\Xi^0)^0$ without changing the set of arrows with that
particular domain. We will say that the subset $\Xi$ is {\em
full\/}\footnote{In the theory of Galois connections, it is
standard to refer to such a set as `closed'. However, this
nomenclature is not used here to avoid confusion with topological
closure.} if $\Xi=(\Xi^0)^0$, and from now on we will write
$(\Xi^0)^0$ as just $\Xi^{00}$. In a similar way, we can show that
if $J\subset \SP_0$ then $J\subset J^{00}:=(J^0)^0$.

Now, for any subset of vectors $\Xi\subset{\cal H}_*$, we have
$\Xi\subset\Xi^{00}$ and hence, in particular,
\begin{equation}
            J^0\subset (J^0)^{00}\label{J0subsetJ000}
\end{equation}
for any subset $J\subset\SP_0$. On the other hand, $J_1\subset
J_2$ implies $J_2^0\subset J_1^0$; hence, in particular, the
relation $J\subset J^{00}$ implies that
\begin{equation}
                (J^{00})^0\subset J^0.\label{J000subsetJ0}
\end{equation}
Putting together \eq{J0subsetJ000} and \eq{J000subsetJ0} we see
that, for any subset $J\subset\SP^0$
\begin{equation}
                    J^0=(J^0)^{00}.
\end{equation}

This means that it is easy to find subsets of non-null vectors
that are full: namely, take the polar, $J^0$, of any subset $J$ of
$\SP_0$ (conversely, any full subset, $\Xi$, of vectors is of the
form $J^0$ for some $J$---just choose $J:=\Xi^0$). In fact, it
would be perfectly reasonable to require from the outset that the
objects in our category $\cal X$ are only {\em full\/} subsets of
vectors.

This is relevant to the remark that, although the approach above
gives genuine generalised truth values of, for example, the type
in \eq{Def:[psi=phi]XQ}, nevertheless there is no obvious physical
significance of the `context' in which such truth values arise:
namely, the subset $\Xi$ of non-null vectors in
\eq{Def:[psi=phi]XQ}. However, if the objects are restricted to be
full subsets of ${\cal H}$, and hence of the form $J^0$ for some
$J\subset\SP_0$, then the context is all those vectors that are
`reducible' with respect to the strings in $J$, which does have
some physical content.

\subsection{A presheaf approach to the normalisation problem}
The basic problem of normalisation is encapsulated in the remark
that if $\hat P$ is a projector such that $\hat P\ket\psi\neq 0$,
then there will invariably be some projectors $\hat Q$ such that
$\hat Q\hat P\ket\psi=0$. The categorial approach in Section
\ref{SubSec:NewCat} is one way of enforcing the non-appearance of
the undesired null vectors under multiplication of projection
operators.

A somewhat different approach is based on the observation that
although, for any given vector $\ket\psi$, $\hat P\ket\psi\neq 0$
does {\em not\/} imply $\hat Q\hat P\ket\psi\neq 0$, the equation
$\hat Q\hat P\ket\psi\neq 0$ {\em does\/} imply that $\hat
P\ket\psi\neq 0$. More generally, if we have a string $Q:=(\hat
Q_q,\hat Q_{q-1},\ldots, \hat Q_1)$ for which $\hat
Q\ket\psi:=\hat Q_q\hat Q_{q-1}\cdots \hat Q_1\ket\psi \neq 0$,
then, necessarily, $\hat Q_{q-1}\hat Q_{q-2}\cdots \hat
Q_1\ket\psi \neq 0$, $\hat Q_{q-2}\hat Q_{q-3}\cdots \hat
Q_1\ket\psi \neq 0$ and so on. Thus although we cannot multiply
projection operators at will, we {\em can\/} `divide' by a
projector in a string for which $\ket\psi$ is reducible. As we
shall now see, this gives another way of handling the
normalisation issue.

The first step is to observe that any monoid $M$ gives rise to a
category, $\tilde M$, whose objects $\tilde M$ are the elements of
$M$, and whose arrows/morphisms are defined by\footnote{It is a
matter of convention which way round the arrows are thought of as
going. Thus it would be equally permissible to define ${\rm
Hom}(m_2,m_1):=\{m\in M\mid m_1=m_2m\}$, and hence ${\rm
Hom}(m_1,m_2):=\{m\in M\mid m_2=m_1m\}$, but we have chosen the
definition in \eq{Def:Homm1m2} as it is the most convenient one
for the application we have in mind. Note that, in
\eq{Def:Homm1m2}, an arrow $m:m_1\rightarrow m_2$ means that $m_2$
is obtained from $m_1$ by `right dividing' $m_1$ by $m$ (not
literally, of course, as $m$ may not be invertible). With the
alternative definition, an arrow $m:m_1\rightarrow m_2$ means that
$m_2$ is obtained from $m_1$ by right multiplying $m_1$ by $m$.}
\begin{equation}
        {\rm Hom}(m_1,m_2):=\{m\in M\mid
        m_1=m_2m\}.\label{Def:Homm1m2}
\end{equation}
The identity arrow $1_m$ is defined as the unit element of $M$ for
all $m\in M$. Note that if $m:m_1\rightarrow m_2$ (so that
$m_1=m_2m$), and $m^\prime:m_2\rightarrow m_3$ (so that
$m_2=m_3m^\prime$) then $m_1=m_2m=m_3m^\prime m$ and hence the
composition $m^\prime\circ m:m_1\rightarrow m_3$ must be defined
as $m^\prime\circ m:= m^\prime m$.

Although $\SP_0$ is only a partial monoid, the general principle
still holds, and we can construct the category $\widetilde{\SP_0}$
whose objects are the elements of $\SP_0$---{\em i.e.}, strings
$Q:=(\hat Q_q,\hat Q_{q-1},\ldots, \hat Q_1)$ for which $\hat
Q:=\hat Q_q\hat Q_{q-1}\cdots \hat Q_1 \neq 0$---and whose arrows
are defined as
\begin{equation}
    {\rm Hom}(Q_1,Q_2):=\{S\in\SP_0\mid Q_1=Q_2\star S\}.
                            \label{Def:HomQ1Q2(a)}
\end{equation}
Note that since the combination law in $\SP_0$ is string
concatenation, there is at most one arrow between any pair of
objects. Hence this particular category is just a partially
ordered set. Note also that if $S_1\in{\rm Hom}(Q_1,Q_2)$ and
$S_2\in{\rm Hom}(Q_2,Q_3)$ then $Q_1=Q_2\star S_1$ and
$Q_2=Q_3\star S_2$, so that $Q_1=(Q_3\star S_2)\star S_1=Q_3\star
(S_2\star S_1)$. Thus the arrow composition in this category is
such that
\begin{equation}
    S_2\circ S_1=S_2\star S_1.           \label{S2circS1}
\end{equation}

The empty string is a terminal object for $\widetilde{\SP_0}$
since, for any object $Q$, we have ${\rm
Hom}(Q,\emptyset):=\{S\in\SP_0\mid Q=S\}=\{Q\}$. Furthermore, if
$Q:=(\hat Q_q,\hat Q_{q-1},\ldots, \hat Q_1)$ is any object, the
unique arrow $Q:Q\rightarrow\emptyset$ factors through a series of
`minimal' arrows that correspond to strings of unit length ({\em
i.e.}, single projection operators):
\begin{eqnarray}
(\hat Q_q,\hat Q_{q-1},\ldots,\hat Q_3,\hat Q_2,\hat Q_1)
&\matrix{(\hat Q_1)\cr \longrightarrow\cr {}}& (\hat Q_q,\hat
Q_{q-1},\ldots, \hat Q_3, \hat Q_2)\longrightarrow \nonumber \\
 &\matrix{(\hat Q_2)\cr \longrightarrow\cr {}} &(\hat
Q_q,\hat Q_{q-1},\ldots, \hat Q_3)\longrightarrow\cdots
\matrix{(\hat Q_{q-1})\cr \longrightarrow\cr {}}(\hat Q_q)
\matrix{(\hat Q_{q})\cr \longrightarrow\cr
{}}\emptyset\hspace{1truecm}\label{ChainArrows}
\end{eqnarray}

Now we discuss state-vector reduction in this context. This
involves introducing the `reduction presheaf' $\R$ on the category
$\widetilde{\SP_0}$. This is defined as follows:
\begin{enumerate}
    \item[(i)] To each object $Q$ in the category $\widetilde{\SP_0}$ we
    associate the space, $\R(Q)$, of vectors that are
    `reducible' with respect to $Q$: {\em i.e.}, vectors $\ket\psi$ on
    which $Q$ acts to give a reduction $\hat Q\ket\psi$ that is
    not the zero vector. Thus\footnote{Equivalently, we could
define $\R(Q)$ to be the set of all {\em rays\/} in $\cal H$ that
are not annihilated by $\hat Q$.}
    \begin{equation}
        \R(Q):=\{\ket\psi\in{\cal H}\mid \hat
        Q\ket\psi\neq0\}    \label{Def:R(Q)}
    \end{equation}
    \item[(ii)] If $S\in{\rm Hom}(Q_1,Q_2)$ is an arrow from $Q_1$
    to $Q_2$ (so that $Q_1=Q_2\star S$) then we define the map
    $\R(S):\R(Q_1)\rightarrow \R(Q_2)$ by
    \begin{equation}
        \R(S)\ket\psi:=\hat S\ket\psi.  \label{Def:R(S)psi}
    \end{equation}
\end{enumerate}
In regard to \eq{Def:R(S)psi}, note that  if $\ket\psi\in\R(Q_1)$
then $\hat Q_1\ket\psi\neq 0$. However, $Q_1=Q_2\star S$ and hence
$\hat Q_1=\hat Q_2\hat S$, and thus $\hat Q_2\hat S\ket\psi\neq
0$. This means precisely that $\hat S\ket\psi\in\R(Q_2)$, and
hence \eq{Def:R(S)psi} does indeed define a map from $\R(Q_1)$ to
$\R(Q_2)$.

Note that if $S_1\in {\rm Hom}(Q_1,Q_2)$ and $S_2\in {\rm
Hom}(Q_2,Q_3)$ then $S_2\circ S_1\in {\rm Hom}(Q_1,Q_3)$ is
defined by \eq{S2circS1} as $S_2\circ S_1=S_2\star S_1$ where, as
we recall, `$\star$' denotes string concatenation. Then, if
$\ket\psi\in\R(Q_1)$, we have
\begin{eqnarray}
    \R(S_2\circ S_1)\ket\psi&=&{\bf R}(S_2\star S_1)\ket\psi
    =\widehat{S_2\star S_1}\ket\psi=\hat S_2\hat S_1\ket\psi
                \nonumber\\
                &=&\R(S_2)\R(S_1)\ket\psi
\end{eqnarray}
so that $\R(S_2\circ S_1)=\R(S_2)\R(S_1)$, as is required for a
presheaf.

Note that, in regard to the chain of arrows in \eq{ChainArrows},
the corresponding actions of the presheaf operators give the chain
of reductions (c.f.\ \eq{StringReductions})
\begin{equation}
\ket\psi\matrix{\R(\hat Q_1)\cr \longrightarrow\cr {}}\hat
Q_1\ket\psi \matrix{\R(\hat Q_{2})\cr \longrightarrow\cr {}} \hat
Q_2\hat Q_1\ket\psi\cdots \matrix{\R(\hat Q_q)\cr
\longrightarrow\cr {}}\hat Q_q\hat Q_{q-1}\cdots \hat Q_1\ket\psi.
\end{equation}

This presheaf can be used to give a contextual, Heyting-algebra
valued generalised truth structure. For example, if $\ket\psi$,
$\ket\phi$ are a pair of vectors in $\R(Q)$ (so that $\hat
Q\ket\psi\neq 0$ and $\hat Q\ket\phi\neq 0$), we provisionally
define\footnote{From a topos perspective, \eq{Def:[psi=phi]Q} is
the characteristic arrow ${\rm eq}_{\R}
:\R\times\R\rightarrow{\bf\Omega}$ of the diagonal subobject
$\triangle:\R\rightarrow\R\times\R$. Here, ${\bf\Omega}$ denotes
the presheaf of sieves on the category $\SP_0$.}
\begin{equation}
    \Big[\ket\psi=\ket\phi\Big]_{\widetilde{\SP_0},Q}:=
            \{S\in{\rm Hom}(Q,\cdot)\mid \hat S\ket\psi=
                            \hat S\ket\phi\}.\label{Def:[psi=phi]QProv}
\end{equation}
Note that if $S\in{\rm Hom}(Q,\cdot)$ then $Q=Q^\prime\star S$ for
some string $Q^\prime$, and therefore, since $\hat Q\ket\psi\neq
0$ and $\hat Q\ket\phi\neq 0$, it follows that $\hat S\ket\psi\neq
0$ and $\hat S\ket\phi\neq 0$ in \eq{Def:[psi=phi]Q} (because
$\widehat{Q^\prime\star S}=\hat Q^\prime\hat S$). We can therefore
normalise the states and replace \eq{Def:[psi=phi]QProv} with (and
referring now to rays in the Hilbert space)
\begin{equation}
    \Big[[\ket\psi]=[\ket\phi]\Big]_{\widetilde{\SP_0},Q}:=
            \left\{S\in{\rm Hom}(Q,\cdot)\mid\exists z\in\mathC,
|z|=1, {\hat S\ket\psi\over\|\hat S\ket\psi}= z{\hat
S\ket\phi\over\|\hat S\ket\phi}\right\}\label{Def:[psi=phi]Q}
\end{equation}
which, of course, is not equivalent to \eq{Def:[psi=phi]QProv}.

The right hand side of \eq{Def:[psi=phi]Q} is {\em contextual\/}
in the sense that it depends on the object $Q$ that is chosen at
which to affirm the statement ``$[\ket\psi]=[\ket\phi]$''. There
could be other spaces $\R(Q^\prime)$ to which both $\ket\psi$ and
$\ket\phi$ belong, and the truth value in the context $Q^\prime$,
namely
$\Big[[\ket\psi]=[\ket\phi]\Big]_{\widetilde{\SP_0},Q^\prime}$,
would not be the same as that in the context $Q$.

The logical structure of these contextual truth values arises
because the right hand side of \eq{Def:[psi=phi]Q} is a {\em
sieve\/} of arrows on $Q$, and hence an element of the Heyting
algebra of all such sieves on $Q$. This is how generalised truth
values arise in the present approach. Note that if $K\subset\cal
H$ is a subset of vectors, all of which are $Q$-reducible (so that
none of them are annihilated by $\hat Q$) then we can define the
valuation
\begin{equation}
\Big[\ket\psi\in K\Big]_{\widetilde{\SP_0},Q}:= \{S\in{\rm
Hom}(Q,\cdot)\mid \hat S\ket\psi\in\hat SK\} \label{Def:[psiinK]Q}
\end{equation}
which gives a contextual, generalised measure of the extent to
which the vector $\ket\psi$ (viewed as a member of $\R(Q)$),
resp.\ the associated ray $[\ket\psi]$, belongs to, resp\ is a
subspace of, the subspace $K\subset\R(Q)$. In particular, if
$K:={\cal H}_{A\in\Delta}$, we arrive at the generalised
valuation\footnote{From a topos perspective, the right hand side
of \eq{Def:[psiinK]Q} is the `evaluation arrow' ${\rm
eval}_{\R}:\R\times P\R\rightarrow{\bf \Omega}$. This is the topos
equivalent of the fact that, in normal set theory, if $J\subset X$
and if $x\in X$, the pair $(x,J)\in X\times PX$ can be mapped to
the value $1 \in \{0,1\}$ if $x\in J$, and to $0\in\{0,1\}$ if
$x\not\in J$.}
\begin{equation}
    V^{[\ket\psi]}(A\in\Delta)_{\widetilde{\SP_0},Q}:=\{S\in{\rm Hom}(Q,\cdot)\mid \hat
S\ket\psi\in\hat S{\cal H}_{A\in\Delta}\}
    \label{Def:VpsiAinDeltaSP0Q}
\end{equation}
which is a sieve at $Q$. We thereby obtain a new candidate for a
generalised truth value for the proposition ``$A\in\Delta$'' in
the context $Q$ when the state is $\ket\psi$ (or, equivalently,
the ray $[\ket\psi]$).

\section{Conclusion}
This paper is a contribution to the long-standing question of
whether the standard quantum formalism  can be given an
interpretation that does not involve measurement as a fundamental
category. This is essential in quantum cosmology, and it is a very
non-trivial problem. Of course, it is quite possible that the
quantum formalism itself needs changing in the cosmological
context, but the working assumption here is that this is not the
case, and that we must therefore strive to give a `neo-realist'
interpretation to standard quantum theory.

In the earlier series of papers by the author and collaborators it
was shown how topos theory could be used to give a generalised
truth value to the propositions in a quantum theory. The topos
concerned involved presheaves over a variety of different
categories, including the category of self-adjoint operators, the
category of Boolean subalgebras of the lattice of projectors, and
the category of abelian von Neumann algebras.

In the present paper we have concentrated instead on the uses of
the topos of $M$-sets for various monoids $M$. We showed that our
earlier results in classical physics can be recovered using the
monoid $C^\infty(\mathR,\mathR)$, and that our earlier results in
quantum physics can be recovered using the monoid
$M(\mathR,\mathR)$.

Then we considered possible applications of the monoid $L({\cal
H})$ of all bounded operators on the Hilbert space $\cal H$ of the
quantum theory. This led rather naturally to thinking about the
monoid of strings of projection operators, and hence ultimately to
the production of a new generalised valuation in quantum theory
whose truth values are determined by what would be state-vector
reductions in the standard instrumentalist interpretation.

If we are not worried about the normalisation issue, then the
final result is \eq{Def:VpsiAinDeltaSP} (or
\eq{Def:VrhoAinDeltaSP} for a density matrix state $\hat\rho$).
This is a {\em bona fide\/} alternative to the valuation
\eq{Def:VpsiAinDelta2} of our earlier work. If the normalisation
problem is of concern, then more sophisticated ideas are needed,
two of which are discussed in the present paper. This leads to the
generalised valuations in \eq{Def:[VpsiAinDeltaXQ} and
\eq{Def:VpsiAinDeltaSP0Q} whose values lie in sieves over the
chosen context/object $\Xi$ and $Q$ respectively. These results
have obvious extensions to the situation where the state is a
density matrix.

It should be emphasised that the material in the present paper
represents only a preliminary investigation of the application of
$M$-sets to quantum theory, and much work remains to be done. In
particular, it is important to see to what extent the
probabilistic predictions in standard quantum theory can be {\em
recovered\/} from the logical values of the generalised valuations
we have discussed above. Ideally, one would like to recover {\em
all\/} the standard probabilistic predictions, so that the logical
structure alone is sufficient to encapsulate the generalised
ontology that is inherent in neo-realist interpretations of the
present type. Hopefully, this will be the subject of a later
paper.

Another potential application of the monoid of strings of
projectors is to consistent history theory in which products of
projectors play a fundamental role \cite{Gri84} \cite{Omn88a}
\cite{GH90b} \cite{Isham94}; one early attempt to discuss
consistent history theory in topos language is \cite{Isham97}.
There are also strong links to the much earlier work on Baer-$*$
rings in quantum logic \cite{Pool75} as well as work on the use of
Galois connections in quantum theory \cite{Butterfield93}.

On the other hand, whilst I was completing this paper, a preprint
appeared very recently with interesting overlaps with some of the
ideas above \cite{LEG06}. This paper deals with an abstract
`algebra of measurements' whose basic ingredient is a monoid of
functions from a space $X$ to itself. In particular, what these
authors call `cumulativity' is related to the ideas above about
using left ideals in $LM$, or sieves. In general, this interesting
approach can clearly be integrated into the discussion of the
present paper. These topics all deserve further study.

\section*{Acknowledgements}
I am most grateful to Jeremy Butterfield for a critical reading of
the draft of this paper. I would also like to thank Andreas
D\"oring for stimulating discussions on the use of topos theory in
physics.

\end{document}